\documentclass{elsart}
\usepackage{amssymb}
\usepackage{amsmath}
\usepackage{epsfig}
\usepackage{psfrag}
\usepackage{subfigure}

\newcommand{\partt}[1]{\frac{\partial {#1}}{\partial t}}
\newcommand{\delsq}{\nabla^2}
\newcommand{\R}{\mathbb{R}}
\newcommand{\Z}{\mathbb{Z}}
\newcommand{\N}{\mathbb{N}}
\newcommand{\Dex}[1]{\Delta\vec{x}_{#1}}

\newcommand{\ie}{i.e.\ }
\newcommand{\rar}{\rightarrow}
\renewcommand{\vec}[1]{\mathbf{#1}}

\begin{document}
\begin{frontmatter}

\title{Spatial and temporal feedback control of traveling wave
   solutions of the two-dimensional complex Ginzburg--Landau
   equation.}
\author{Claire M.\ Postlethwaite} and
\author{Mary Silber}
\address{Department of Engineering Sciences and Applied Mathematics\\
   Northwestern University \\ Evanston, IL 60208 USA}

\begin{abstract}
Previous work has shown that Benjamin--Feir unstable traveling waves
of the complex Ginzburg--Landau equation (CGLE) in two spatial
dimensions cannot be stabilized using a particular time-delayed
feedback control mechanism known as `time-delay
autosynchronisation'. In this paper, we show that the addition of
similar \emph{spatial} feedback terms can be used to stabilize such
waves. This type of feedback is a generalization of the time-delay
method of Pyragas (\emph{Phys. Letts. A} {\bf 170}, 1992) and has been
previously used to stabilize waves in the one-dimensional CGLE by
Montgomery and Silber (\emph{Nonlinearity} {\bf 17}, 2004). We
consider two cases in which the feedback contains either one or two
spatial terms. We focus on how the spatial terms may be chosen to
select the direction of travel of the plane waves. Numerical linear
stability calculations demonstrate the results of our analysis.
\end{abstract}
\end{frontmatter}

\section{Introduction}

During the past two decades considerable progress has been made in our
understanding of the spontaneous emergence of patterns in
spatially-extended non-equilibrium systems. The existence of simple
spatial or spatio-temporal patterns has been rigorously established on
the basis of equivariant bifurcation theory (see~\cite{GSS88} and
references therein). However, these simple patterns are often unstable
in a given system, which evolves instead to a state of spatio-temporal
chaos. Patterns that result from a symmetry-breaking Hopf bifurcation
seem to be especially vulnerable to instability.  A current challenge
in pattern-formation research is to develop control schemes that
stabilize the simple patterned states, so that a desired, otherwise
unstable, solution may be realized.  This paper extends current
research on feedback control of oscillatory patterns, focusing on
traveling wave solutions of the two-dimensional complex
Ginzburg--Landau equation. One of the goals of this line of research
is to develop control algorithms that exploit the underlying
symmetries of a targeted pattern in order to stabilize it in a
non-invasive fashion. That is, if the pattern is invariant under some
spatial, temporal or spatio-temporal transformation, then it may be
possible to construct a feedback based on this transformation which
vanishes when the target pattern is achieved. For example, if the
pattern has the property that $A(\vec{x},t)=A(R\vec{x},t-\tau)$ where
$R\vec{x}$ is some Euclidean transformation of the spatial variable,
and $\tau\geq 0$ is a possible time delay, then a feedback which is
proportional to $A(R\vec{x},t-\tau)-A(\vec{x},t)$ will be non-invasive.

Ott, Grebogi and Yorke~\cite{OGY90} were the first to develop a
control algorithm that suppresses chaos in favor of a simple unstable
periodic orbit (UPO) in low-degree-of-freedom chaotic systems. In this
approach, small perturbations are applied to a system parameter in
order to keep the system close to the UPO. However, this method
requires constant monitoring of the system, and also may not be
effective in rapidly evolving systems.

Pyragas~\cite{Pyr92} introduced a second approach, sometimes called
`time-delayed autosynchronisation' (TDAS), in which the feedback is
proportional to the difference between the current and a past state of
the system. That is, the feedback is $F=\gamma(A(t)-A(t-\Delta t))$
where $\Delta t$ is the period of the desired UPO, and $A(t)$ is some
state variable. This method has a number of attractive
properties. First, the UPO of the original system will also be a
solution of the system with feedback, and so control may be achieved
in a non-invasive manner. Second, the only information required a
priori is the period of the desired UPO.  The method has been
implemented successfully in a variety of laboratory experiments on
electronic \cite{PT93,GSCS94}, laser \cite{BDG94}, plasma
\cite{PBA96,MKPAPB97,FSK02}, mechanical~\cite{FEA06} and chemical
systems \cite{SBFHLM93,LFS95,PMRNKG,KKG06}; more examples can be found
in a recent review by Pyragas~\cite{Pyr06}.

An extension of TDAS proposed by Socolar \etal~\cite{SSG94}
incorporates information at many previous times and is known as
`extended time-delayed autosynchronisation' (ETDAS). This method was
used by Bleich and Socolar~\cite{BS96b} to stabilize unstable
traveling wave solutions of the complex Ginzburg--Landau equation
(CGLE) in one spatial dimension. For spatially-extended
pattern-forming systems, other proposed modifications of the
time-delay autosynchronization scheme of Pyragas include \emph{global}
feedback control~\cite{BM04,GN06,BB06}, where the magnitude of the
feedback depends on some spatial average, or maximum, of a quantity at
an earlier time. Yet others take into account the spatial periodicity,
as well as temporal periodicity, of the targeted pattern. For
instance, Lu, Yu and Harrison~\cite{LYH96} used numerical simulations
of the two-dimensional Maxwell--Bloch equations describing a
three-level laser system to demonstrate that spatio-temporal chaos
could be suppressed by applying a linear combination of time-delay
feedback and an analogous spatially-translated feedback term of the
form:
\begin{equation}
F_s=\rho\{[E(x+x_0,y,t)
-E(x,y,t)]+[E(x,y+y_0,t)-E(x,y,t)]\}.
\nonumber
\end{equation}
Here ${\bf d}_1=(x_0,0)$ and ${\bf d}_2=(0,y_0)$ are translation
vectors associated with the feedback; the feedback is non-invasive
when the resulting pattern is periodic in each of the ${\bf  
d}_j$-directions,
with spatial periods $|{\bf d}_j|$. Their numerical investigations
reveal that these two spatial controls somehow select the direction of  
travel
of the stabilized plane wave. This type of feedback, with a combination  
of temporal and
spatial terms, was extensively studied by Montgomery and
Silber~\cite{MS04} in the context of  the one-dimensional CGLE.

The aim of our paper is to extend the results of~\cite{MS04} to the
CGLE in two spatial dimensions. This problem is interesting since
Harrington and Socolar~\cite{HS01} have shown that unstable traveling
waves in the 2D CGLE {\it cannot} be stabilized using only temporal
feedback, due to the presence of torsion-free modes (that is, modes
which have purely real Floquet multipliers). We circumvent this
difficulty by using a combination of temporal and spatial terms, as
in~\cite{MS04}, and additionally consider the effect of having either
one or two spatially shifted terms in order to steer the direction of
the traveling waves in the plane.

Our analysis is based on a linear stability analysis of the traveling
wave solutions of the CGLE. The analysis leads to a system of delay
differential equations (see~\cite{Diek,Driver} for more on DDEs).  We
find stability boundaries by searching for critical curves/surfaces of
Hopf bifurcations within the system of DDEs.  The analysis is made
possible because the CGLE admits an exact family of solutions in the
form of a traveling plane waves $R\e^{i\vec{k}\cdot\vec{x}+i\omega
t}$, parameterized by the wave vector $\vec{k}$. For this problem the
dispersion relation that relates $|\vec{k}|$ and $\omega$ is known
precisely.

We expect that our methods could equally well be applied to other
spatially extended systems for which traveling plane waves exist. Such
solutions arise for instance when a spatially-uniform solution
undergoes a Hopf bifurcation.  However, in many instances, the form of
the wave and the dispersion relation are unknown, or known only
approximately. In this case, we propose that just one spatial term be
incorporated in the feedback, and consider target waves that travel in
a direction relative to this feedback term so that both the temporal
and spatial feedback terms vanish.  In this way the wavenumber and
frequency of the plane wave can be matched even when the dispersion
relation is unknown, although the direction of travel must be left
free.

This paper is organized as follows. In Section~\ref{sec:setup} we
review some properties of the CGLE and the feedback we are applying,
and in Section~\ref{sec:prev} we describe previously known results
regarding these systems. In Section~\ref{sec:spfe} we discuss the
effects of additional spatial feedback and how the spatial shifts must
be chosen in order to stabilize the wave. Section~\ref{sec:anal}
contains a geometric way to think about the results of~\cite{MS04}
and in Section~\ref{sec:num} we provide a numerical example to
demonstrate our results. Section~\ref{sec:conc} concludes.

\section{Problem setup}
\label{sec:setup}

\subsection{The complex Ginzburg--Landau equation}

In this section we review some properties of traveling wave solutions
of the complex Ginzburg--Landau equation (CGLE). The CGLE is an
amplitude equation for some disturbance $A(\vec{x},t)$ in a spatially
extended system near the onset of a Hopf bifurcation that sets in with
zero wavenumber, {\it i.e.} a spatially--uniform oscillatory
instability. After appropriate rescalings and in a frame that is
oscillating at the Hopf frequency, it can be written generically as:
\begin{equation}\label{eq:CGLE}
\partt{A}=A+(1+ib_1)\delsq A-(b_3-i)|A|^2A,
\end{equation}
where $b_1$ and $b_3$ are real parameters. We consider only the
  situation just after a supercritical Hopf bifurcation, so the
  coefficient of the linear term is positive (and has been rescaled to
  equal unity) and the parameter $b_3$ is assumed to be positive. In
  this paper, we consider the spatial extent of the problem to be
  two-dimensional, so $\vec{x}=(x,y)$.  We consider traveling wave
  solutions of the form
\begin{equation}\label{eq:TWsoln}
A_{\mathrm{TW}}=R\e^{i\vec{k}\cdot\vec{x}+i\omega t},
\end{equation}
where the amplitude $R$ and frequency $\omega$ are determined by the
wavenumber $k=|\vec{k}|$ through the relations
\begin{equation}\label{eq:Rw}
R^2=\frac{1-k^2}{b_3},\qquad \omega=R^2-b_1k^2.\end{equation}
Since $b_3>0$ it follows from~\eqref{eq:Rw}
that $k<1$. We also assume that $\vec{k}\neq\vec{0}$.

In the Benjamin--Feir unstable regime ($b_1>b_3>0$) all solutions of
the form~\eqref{eq:TWsoln} are unstable. In this paper we investigate  
how the addition
of feedback terms to~\eqref{eq:CGLE} affects the linear stability of
these solutions in this regime. Specifically we consider
\begin{equation}\label{eq:CGLEF}
\partt{A}=A+(1+ib_1)\delsq A-(b_3-i)|A|^2A+F,
\end{equation}
where $F$ is a feedback term given by
\begin{equation}\label{eq:feed}
F=\gamma[A(\vec{x},t)-A(\vec{x},t-\Delta
   t)]+\sum_{j=1}^N\rho_j[A(\vec{x}+\Delta
   \vec{x}_j,t)-A(\vec{x},t)].
\end{equation}
The delay parameter $\Delta t$ is positive, and we restrict our
analysis to the case where $\gamma$ (the gain) and $\rho_j$ are
real. We consider the two cases $N=1$ and $N=2$, that is, the feedback
$F$ can contain either one or two spatially-shifted feedback terms.

In this paper, we study the stability of traveling waves for which the  
feedback terms vanish. We
choose the spatial and temporal shifts $\Dex{j}$ and $\Delta t$ such  
that
\begin{equation}\label{eq:shifts_van} \Delta t |\omega|= 2\pi \qquad  
\Delta\vec{x}_j\cdot\vec{k}=2\pi n_j,\quad  n_j\in\Z,\  
j=1,\dots,N\end{equation}
for a specific targeted traveling wave of the form~\eqref{eq:TWsoln},
with wavevector $\vec{k}=(k_x,k_y)$ and frequency $\omega$.

There may be many other solutions to~\eqref{eq:CGLEF} for which the  
feedback term does not
vanish. We do not study these solutions in this paper - our analysis
is a purely local stability analysis assuming we are already close to  
the targeted wave.

Since we know the dispersion relation~\eqref{eq:Rw} relating the wave  
frequency
$\omega$ and the wavenumber $k=|\vec{k}|$, the
spatial and temporal shifts are related and cannot be chosen
independently. We now consider how the shifts must be chosen in the
two cases $N=2$ and $N=1$.

Note that if we were to choose $N>2$, there would be some relations  
between the $\Dex{j}$ in order that they satisfy~\eqref{eq:shifts_van}.  
For instance, for $N=3$, we would need $\Dex{3}=n_1\Dex{2}+n_2\Dex{2}$  
for some integers $n_1$ and $n_2$. This is because we are in two  
spatial dimensions. We do not consider the cases of higher $N$ here,  
but it may be that using additional shifted terms can increase the  
region of stability of the traveling waves, much in the same way as  
ETDAS~\cite{BS96b}.

For $N=2$, we insist that $\Dex{1}$ and $\Dex{2}$ are not parallel
($\Dex{1}\times\Dex{2}\neq 0$). Then for specified $\Dex{1}$ and
$\Dex{2}$ there is a two-dimensional dual lattice of possible target
wavevectors $\vec{k}$. The lattice generators $\vec{k}_1$ and
$\vec{k}_2$ satisfy
\begin{equation}\label{eq:klatt}
|\vec{k}_j|=\frac{2\pi|\Dex{j}|}{|\Dex{1}\times\Dex{2}|},\qquad
\vec{k}_j\cdot\Dex{j}=0,\quad j=1,2.
\end{equation}

The frequency of the desired wave is specified by the choice of $\Delta  
t$,
which must be consistent with the choice of $\vec{k}$, since the
frequency and wavenumber are related {\it a priori} by the dispersion
relation~\eqref{eq:Rw}. (This `overspecification' of the targeted wave
is analogous to the one-dimensional case of~\cite{MS04}, where
one temporal and one spatial term select $\omega$ and $k$ which
are already related by the dispersion relation.)
Note that if $\vec{k}$ lies on the lattice, so
also does $\vec{-k}$, and thus the direction (right/left) of travel of
the wave is not selected.

If the same procedure were repeated for a different amplitude equation
for which the dispersion relation were not known, then in the
`overspecified' $N=2$ case there could be a discrepancy between the
temporal and spatial feedback terms, and it may not be possible for
both to vanish. However, in the case $N=1$, there is more freedom, as
we explain below.

For $N=1$ it is not possible to choose $\Dex{1}$ such that the
spatially-shifted term targets a single traveling wave. That
is, for a given $n_1$, there is a continuum of wavevectors $\vec{k}$  
for which
$\vec{k}\cdot\Dex{1}=2\pi n_1$ and the spatial feedback term vanishes. The
possible target $\vec{k}$ for a given $\Dex{1}$ are indicated by the  
bold lines in
Figure~\ref{fig:1dx}, in section~\ref{sec:spfe}.
However, the choice of $\Delta t$ selects
some $\omega$, and hence $|\vec{k}|$ is also selected, by the  
dispersion relation. The direction
of $\vec{k}$ (again, up to a sign) is selected by the spatially shifted  
term, and so a
single traveling wave can be targeted. This is explained in more detail  
in Section~\ref{sec:N1}.

Note that $F$ can only improve the stability of the traveling
waves if $\gamma< 0$ and likewise $\rho_j> 0$. The reasons for this are  
explained in
detail in~\cite{MS04} and we will only consider this parameter regime.

\subsection{Stability analysis}

The linear stability of traveling waves~\eqref{eq:TWsoln}
is calculated by considering
the effect of small amplitude perturbations,
for some perturbation wavevector $\vec{q}=(q_x,q_y)$. We write
\begin{equation}\label{eq:perts}
A=R\e^{i\vec{k}\cdot\vec{x}+i\omega    
t}(1+a_+(t)\e^{i\vec{q}\cdot\vec{x}}+a_-(t)\e^{-i\vec{q}\cdot\vec{x}}),
\end{equation}
substitute into~\eqref{eq:CGLEF} and linearize in $a_+$ and $a_-$.
This results in a system of delay differential equations for
$a_+$ and the complex conjugate of $a_-$ (denoted by $a_-^*$):
\begin{equation}\label{eq:dde}\frac{d}{dt}\begin{pmatrix}a_+(t) \\
     a_-^*(t) \end{pmatrix}=J\begin{pmatrix}a_+(t) \\ a_-^*(t)
   \end{pmatrix}+\gamma\left[\begin{pmatrix}a_+(t) \\
     a_-^*(t) \end{pmatrix}-\begin{pmatrix}a_+(t-\Delta t) \\  
a_-^*(t-\Delta
   t) \end{pmatrix}\right],\end{equation}
where
\begin{equation}\label{eq:J} J=\begin{pmatrix}
     -c_1 q^2-c_2 R^2 & -c_2 R^2 \\
-c_2^*R^2 &  -c_1^* q^2-c_2^*R^2
\end{pmatrix}+ 2\vec{k}\cdot\vec{q}\begin{pmatrix}-c_1 & 0 \\ 0
     & c_1^*\end{pmatrix}
+\sum_{i=1}^N\rho_j(\e^{i\vec{q}\cdot\Dex{j}}-1)\begin{pmatrix} 1 & 0  
\\ 0 & 1 \end{pmatrix}
\end{equation}
with $c_1=1+ib_1$, $c_2=-i+b_3$ and $q=|\vec{q}|$.

In the following sections we review previous results concerning
feedback control in equations similar to~\eqref{eq:dde} and then  
present our new results.

\section{Previous results}
\label{sec:prev}

Our stability analysis builds on results of Harrington and
Socolar~\cite{HS01}, Nakajima~\cite{Nak97} and Montgomery and
Silber~\cite{MS04}. For completeness, in this section we give a summary  
of the results
of~\cite{HS01} and~\cite{Nak97}, and also of previously known results
concerning some instabilities of the CGLE without feedback. We discuss
details of~\cite{MS04} in Section~\ref{sec:anal}.

\subsection{Instabilities with no feedback terms}
\label{sec:nofeed}

In the absence of any feedback terms (\ie $\gamma=\rho_j=0$), traveling  
waves are stable to
perturbations with wavevector $\vec{q}$ if both eigenvalues of the
Jacobian~\eqref{eq:J} have negative real part.

The Benjamin--Feir (B--F) instability is a long wavelength
instability, that is, perturbations with $q=|\vec{q}|$ sufficiently
small will grow. In the limit $q\ll 1$, a simple calculation shows
that one eigenvalue of $J$ always has negative real part, and the
other has real part equal to:
\[\left(\frac{b_1}{b_3}-1\right)q^2+
\frac{2}{b_3R^2}\left(1+\frac{1}{b_3^2}\right)(\vec{k}\cdot\vec{q})^2+O( 
q^3).\]
Hence, if $b_1>b_3>0$ (the B--F unstable regime), then this is
positive for all $\vec{k}$, and long wavelength
perturbations will grow.

A second type of instability important to our subsequent analysis, is
associated with perturbations
with $\vec{k}\cdot\vec{q}=0$ ($q$ no
longer necessarily small). It can be shown~\cite{HS01} that these perturbations grow for
small enough $q$, that is, those with
\begin{equation}\label{eq:qcr}
q^2<q_{\mathrm{cr}}^2=\frac{2R^2(b_1-b_3)}{1+b_1^2}.
\end{equation}
Perturbations with $\vec{k}\cdot\vec{q}=0$ and $q>q_{\mathrm{cr}}$  
decay.

When $q$ is sufficiently large, the eigenvalues of $J\sim -q^2$, so all  
sufficiently shortwave perturbations will decay. However, there may be  
other regions of unstable $\vec{q}$ in addition to those described  
above.

\subsection{A result of Harrington and Socolar}

We now summarize a result from Harrington and Socolar~\cite{HS01}
concerning the stabilization of plane waves in the 2D CGLE with only
temporal feedback. Their result relies on a proof of
Nakajima~\cite{Nak97} which has recently~\cite{Fie06, Jus07} been
shown to be incorrect. However, as we discuss below, the result of
Nakajima does still hold in some cases, and the results of~\cite{HS01}
are still correct. 

The claim of Nakajima is as follows. Suppose an ODE
\begin{equation}\label{eq:ode}\dot{x}=f(x(t),t),\qquad x\in  
\R^n,\end{equation}
contains an unstable periodic orbit, $x^*(t)$, with period $T$. Feedback is added to~\eqref{eq:ode}
so it is now of the form
\begin{equation}\label{eq:ode_feed}\dot{x}(t)=f(x(t),t)+K(x(t)-x(t- 
T)),\end{equation}
where $K$ is an $n\times n$ gain matrix. Then if the linearization of~\eqref{eq:ode} about the orbit $x^*(t)$
has an odd number of positive, real
unstable Floquet multipliers then there is no value
of $K$ for which $x^*(t)$ is a stable solution of~\eqref{eq:ode_feed}.

The flaw in the proof of Nakajima is that the neutral Floquet
multiplier associated with perturbations in the direction of the periodic orbit was neglected.

However, in both the example of Harrington and Socolar, and in our work,
the analysis of the stability of the traveling wave has
been reduced to the study of the stability of a fixed point (namely,
the zero solution to the linear DDE~\eqref{eq:dde}). The result of
Nakajima \emph{can} be applied to hyperbolic fixed points, since they
do not have a trivial Floquet multiplier, and so the results
of~\cite{HS01} still stand. (A similar result for fixed points in
maps, rather than flows, is given by Ushio~\cite{Ush96}.)
In our examples, the perturbation along the periodic orbit which
yields a neutral Floquet multiplier is associated with a perturbation 
with wavevector $\vec{q}=\vec{0}$. It is not this type of perturbation
we are concerned with in the following.

The result of Harrington and Socolar is summarized as follows. Consider~\eqref{eq:dde} with no
spatial feedback, that is, $\rho_j=0$. In 2D
there always exists an unstable perturbation wavenumber $\vec{q}$
which satisfies $\vec{k}\cdot\vec{q}=0$. In this situation, the Jacobian $J$ is equal to
\[\begin{pmatrix}
     -c_1 q^2-c_2 R^2 & -c_2 R^2 \\
-c_2^*R^2 &  -c_1^* q^2-c_2^*R^2
\end{pmatrix}\]
where $\mathrm{Re}(c_1),\mathrm{Re}(c_2)>0$. If $0<q<q_{cr}$, then $J$  
has real trace
and real, negative determinant, hence has real eigenvalues of opposite
sign.

Note that in only one spatial dimension, we can
only have $\vec{k}\cdot\vec{q}=\vec{0}$ if either $\vec{q}=\vec{0}$ or  
$\vec{k}=\vec{0}$, so the same result
does not apply. The `flat' perturbations with
$\vec{q}=\vec{0}$ do not cause instabilities for $\gamma\leq 0$, which is the  
parameter range we are interested in.

\section{Additional spatial feedback}
\label{sec:spfe}

In this section we extend the results described above
to include the effects of additional spatial feedback, with either one
or two spatial terms. This section is organized as follows. First, we
show that there are some perturbation wavevectors $\vec{q}$ which are
unaffected by the spatial feedback. We then discuss how the
displacements $\Dex{j}$ must be chosen to target a particular
wavevector $\vec{k}$, for both the cases $N=1$ and $N=2$.

\subsection{Perturbations unaffected by spatial feedback}
\label{sec:unaffect}

Consider the Jacobian matrix $J$ given in~\eqref{eq:J}. If
the perturbation wavevector $\vec{q}$ satisfies
$\vec{q}\cdot\Dex{j}=2\pi m_j$, for some $m_j\in\Z$, then $J$ is the  
same
as if $\rho_j=0$. Hence, these perturbations $\vec{q}$ are not affected  
by the
spatial feedback. Combining this with the results of Harrington and
Socolar~\cite{HS01} described in the previous section, if these
$\vec{q}$ are unstable for the system with no feedback, and
additionally satisfy $\vec{k}\cdot\vec{q}=0$,
then the traveling waves cannot be stabilized for any values of  
$\rho_j$ and $\gamma$.
That is, if there is a $\vec{q}\neq\vec{0}$ which satisfies
\begin{equation}\label{eq:dangq}
\vec{k}\cdot\vec{q}=0,\quad\vec{q}\cdot\Dex{j}=2\pi n_j,\quad n_j\in\Z  
\quad
\mathrm{and} \quad|\vec{q}|<q_{cr},\end{equation}
then the traveling waves cannot be stabilized.

If there are no such $\vec{q}$, then it is possible that there exists
some choice of $\rho_j$ and $\gamma$ for which the traveling wave is  
stable.

\subsection{Choice of $\Dex{j}$}

Recall that in order for the spatial feedback to vanish at the targeted  
wave
solution, we require that
\begin{equation}
\vec{k}\cdot\Dex{j}=2\pi n_j,\quad n_j\in\Z.
\end{equation}
In one spatial dimension, this gives a unique choice of $\Dex{j}$ (for
say, $n_j=1$) for a given $\vec{k}$, but in two spatial dimensions
this is not the case. We now discuss this further.

\subsubsection{The case $N=1$}
\label{sec:N1}

\begin{figure}
\begin{center}
\hspace{-4cm}
{\psfrag{kx}{$k_x$}
\psfrag{ky}{$k_y$}
\psfrag{kp}{$\vec{k}_p$}
\psfrag{k}{$\vec{k}$}
\psfrag{q}{$\vec{q}$}
\psfrag{0}{$\vec{k}\cdot\Dex{1}=0$}
\psfrag{m2}{$\vec{k}\cdot\Dex{1}=-2\pi$}
\psfrag{m4}{$\vec{k}\cdot\Dex{1}=-4\pi$}
\psfrag{p2}{$\vec{k}\cdot\Dex{1}=2\pi$}
\psfrag{p4}{$\vec{k}\cdot\Dex{1}=4\pi$}
\psfrag{ktext}{$k^2$ as given by dispersion relation}
\psfrag{qtext}{\shortstack{Example $\vec{q}$ which \\  satisfies  
$\vec{k}\cdot\vec{q}=0$}}
\epsfig{file=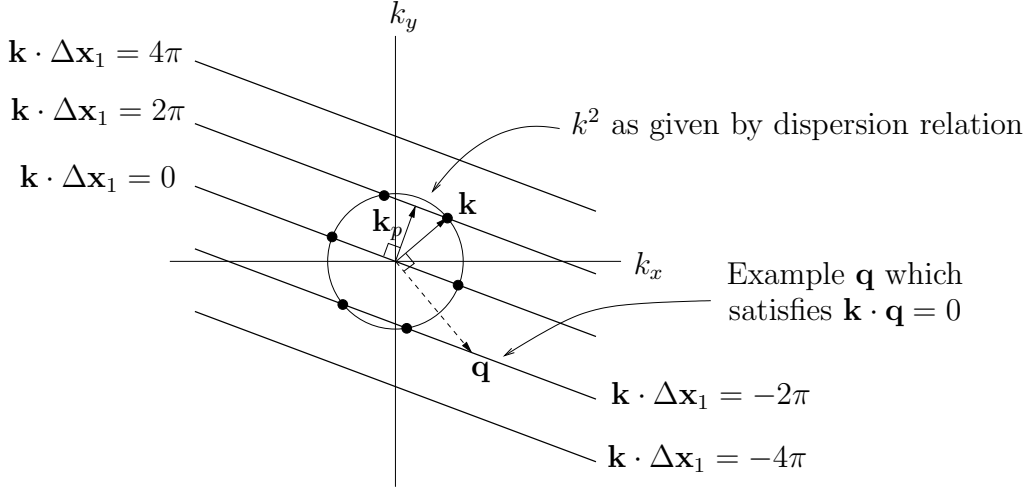, width=10cm}}
\end{center}
\caption{\label{fig:1dx} The bold lines indicate those points in
   $\vec{k}$-space which satisfy $\vec{k}\cdot\Dex{1}=2\pi
   n_1$ ($n_1\in\Z$). The vector $\vec{k}_p$ is parallel to $\Dex{1}$.  
The
   circle indicates the wavevectors $\vec{k}$ for which the temporal
   feedback term vanishes with $|\omega|\Delta t=2\pi$.
 The intersections of the circle and
   the lines (marked by dots) are the possible target $\vec{k}$. As the
   direction of $\vec{k}$ gets closer to $\vec{k}_p$, either by changing
   the size of the circle, or altering the angle of $\Dex{1}$, the
   magnitude of the $\vec{q}$ which satisfy $\vec{k}\cdot\vec{q}=0$ and
   also lie on the bold lines increases.
}
\end{figure}

For $N=1$ there is a continuum of wavevectors $\vec{k}$ for which
$\vec{k}\cdot\Dex{1}=2\pi n_1$ and the feedback term vanishes. The
possible $\vec{k}$ are indicated by the bold straight lines in
Figure~\ref{fig:1dx}.
However, isolated $\vec{k}$ can be picked out from this continuum by  
the
temporal feedback term (which fixes $\Delta t$ and hence $\omega$) and  
the dispersion relation. The dispersion relation
fixes $k^2$ for a given $\omega$ and so gives a circle of possible
wavevectors. If we let $|\omega|\Delta t=2\pi n$ for some integer  
$n>1$, then this will give a family of circles, of decreasing $k$  as  
$n$ increases (since $k$ must decrease as $\omega$ increases), however,  
we only consider $n=1$.
The intersection of this circle and the straight lines in
Figure~\ref{fig:1dx} give the possible $\vec{k}$ for which the
feedback term vanishes. In this example, there are three possible  
$\vec{k}$ (up to a sign). As we explain below, the $\vec{k}$ with  
direction closest to that of $\Dex{1}$ will be the easiest to  
stabilize. However, it is certainly possible that more than one of  
these $\vec{k}$ may be simultaneously stabilized.

If the dispersion relation for a system is known, as is the case for
the CGLE, then the location of
this circle is known, and $\Delta t$ and $\Dex{1}$ can be chosen to
target a particular traveling wave. If the dispersion relation is not
known, then the location of the circle is not known, but it still
exists. In this case, we can, say, choose $\Delta t$ to pick the
frequency of the target wave, and also pick some $\Dex{1}$. We will
not be able to choose in advance the wavevector $\vec{k}$ of the
resulting targeted wave, but the possible
$\vec{k}$ for which the feedback vanishes will be isolated in
$\vec{k}$-space. This method provides us with a possible mechanism for
using this stabilization scheme for a system in which the dispersion
relation is not known.

If there is a $\vec{q}$ which satisfies~\eqref{eq:dangq}, it will
also lie on the bold lines in Figure~\ref{fig:1dx}. However, for a
given $\vec{k}$, $\Dex{1}$ can be chosen in such a way that there are
no $\vec{q}$ which satisfy the resonance conditions~\eqref{eq:dangq}.
To see this, consider the situation where
$\Dex{1}$ is chosen to be almost parallel to $\vec{k}$, then the only
$\vec{q}$ perpendicular to $\vec{k}$ which also lie on the bold lines
will have a very large magnitude and hence do not correspond to
wavenumbers of destabilizing perturbations.  Once $\Dex{1}$
is chosen to avoid the resonance conditions~\eqref{eq:dangq} the methods
developed in~\cite{MS04} can then be used to determine the stability of  
the
traveling waves.

By analogy with the one-dimensional case, we might hope that
we are able to stabilize the wave for which $\vec{k}$
and $\Dex{1}$ are parallel (indicated by $\vec{k}_p$ on
Figure~\ref{fig:1dx}).
However, this is not the case since
there always exist unstable longwave perturbations ($q$ small)
which satisfy $\vec{k}_p\cdot\vec{q}=\vec{q}\cdot\Dex{1}=0$, and these
perturbations are
unaffected by the spatial feedback. Hence this plane wave cannot be
stabilized in this fashion.

\subsubsection{The case $N=2$}

\begin{figure}
\begin{center}
{\psfrag{kx}{$k_x$}
\psfrag{ky}{$k_y$}
\psfrag{k1}{}
\epsfig{file=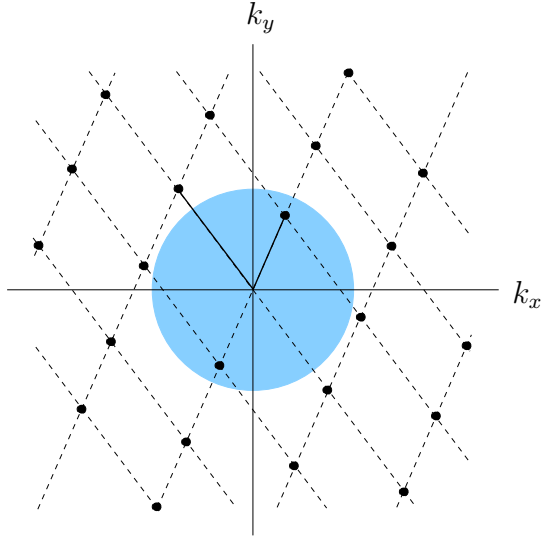, width=7cm}}
\end{center}
\caption{\label{fig:klat} The figure shows a lattice of possible  
$\vec{k}$-vectors
   for which the spatial feedback vanishes in the case $N=2$. The bold  
lines indicate the
   lattice generators. The shaded region shows those $\vec{k}$ with
   $|\vec{k}|<1$. In this example only two non-trivial lattice points lie
   within this disc.}
\end{figure}

For the spatial feedback with two terms to vanish at the targeted wave
solution, $\vec{k}$ must satisfy
\[\vec{k}\cdot\Dex{1}=2\pi n_1,\qquad\vec{k}\cdot\Dex{2}=2\pi
n_2,\quad n_1,n_2\in\Z.\]
This gives a lattice of possible target wavevectors with generators as
given in equation~\eqref{eq:klatt}. An example lattice is shown in  
Figure~\ref{fig:klat}.
Recall that for the CGLE we must have $k=|\vec{k}|<1$, so for a given
target $\vec{k}$, it is possible to choose $\Dex{1}$ and $\Dex{2}$
such that $\vec{k}$ and $-\vec{k}$ are the only non-trivial lattice
points inside the unit circle.  However, for a specific choice of
$\vec{k}$, the choice of $\Dex{1}$ and $\Dex{2}$ resulting in an
appropriate lattice is not unique.

The perturbations $\vec{q}$ which satisfy~\eqref{eq:dangq} will lie on
the same lattice as $\vec{k}$. If $q_{cr}$ is small enough, there will
be no unstable $\vec{q}$ (that is, with $|\vec{q}|<q_{cr}$) which lie  
on the lattice.
Therefore, so long as $\Dex{1}$ and $\Dex{2}$ are chosen
appropriately, the theorem of Nakajima does not apply. The methods
developed in~\cite{MS04} can then be used to determine the regions of
stability for these waves. In the next section we review the approach  
taken
in~\cite{MS04}.

\section{Analysis}
\label{sec:anal}

In this section we discuss a result found in Montgomery and
Silber~\cite{MS04}. They show that for the CGLE in 1D with feedback as
in~\eqref{eq:CGLEF}, the `best' choice of $\gamma$ is $-1/\Delta t$.  
That is, if it is
possible to stabilize a traveling wave for some choice of feedback
parameters, this can certainly be achieved by choosing  
$\gamma=-1/\Delta t$.

The proof given in~\cite{MS04} can be extended to two dimensions. Here
we omit those details and give a briefer, and more intuitive reason as to why their result
holds, although it is not rigorous.

We first reduce the delay differential equation~\eqref{eq:dde} to a
single equation by diagonalizing the matrix $J$ to give
\begin{align}\dot{a}(t)&=\hat{m}_1a(t)+\gamma(a(t)-a(t-\Delta t)) \\
\dot{b}(t)&=\hat{m}_2b(t)+\gamma(b(t)-b(t-\Delta t))
\end{align}
where $a$ and $b$ are appropriate linear combinations of $a_+$ and  
$a^*_-$, and
  \[\hat{m}_k=m_k+\sum_{j=1}^2\rho_j(\e^{i\vec{q}\cdot\Dex{j}}-1),\quad
  k=1,2.\]
  The $m_k$ are the eigenvalues of the Jacobian~\eqref{eq:J} with no  
feedback (that
  is, $\rho_j=\gamma=0$), with $\mathrm{Re}(m_1)>\mathrm{Re}(m_2)$. As  
described in~\cite{MS04}, the $b$ equation does not affect
  the stability boundary since it is associated with the more stable
  eigenvalue. We therefore concentrate on the $a$ equation,
  and write $\hat{m}_1=\alpha(\vec{q})+i\beta(\vec{q})$, where $\alpha$
  and $\beta$ are real functions of $\vec{q}$. Hence, $\alpha(\vec{q})$  
is the largest real
part of the eigenvalues of $J$ when there is no temporal
  feedback. Since the feedback can only improve the stability of the
  traveling waves, instabilities are only possible when  
$\alpha(\vec{q})>0$.

Critical surfaces describing the boundaries between regions of
stable and unstable perturbations have $a(t)=\e^{\lambda t}$ with
$\lambda=i\nu$, $\nu\in\R$ (\ie they are points of Hopf bifurcations
since $\nu\ne 0$).
We can write equations for these surfaces in
$\gamma$-$\vec{q}$ space, parameterized by $\nu$, as:
\begin{align}
\alpha(\vec{q})&=\gamma(\cos\nu-1)\label{eq:cc1}\\
\gamma\Delta t\sin\nu &=\nu+\Delta t\beta(\vec{q}).\label{eq:cc2}
\end{align}
The results of~\cite{MS04} can be used to give an algebraic condition
for the stability of the traveling waves. That is, if there are no
solutions $\nu$, $\vec{q}$ to the equations
\begin{align}
\cos\nu-1&=-\Delta t\alpha(\vec{q})\\
\sin\nu+\nu &=-\Delta t\beta(\vec{q})
\end{align}
then the traveling wave can be stabilized at $\gamma=-1/\Delta t$.

\begin{figure}
\begin{center}
{\psfrag{g}{$\gamma$}
\psfrag{a}{$\alpha(\vec{q})=0$}
\psfrag{q}{$|\vec{q}|$}
\psfrag{dt}{$-\frac{1}{\Delta t}$}
\psfrag{lt}{$\alpha(\vec{q})<0$}
\psfrag{gt}{$\alpha(\vec{q})>0$}
\psfrag{s}{s}
\psfrag{u}{u}
\epsfig{file=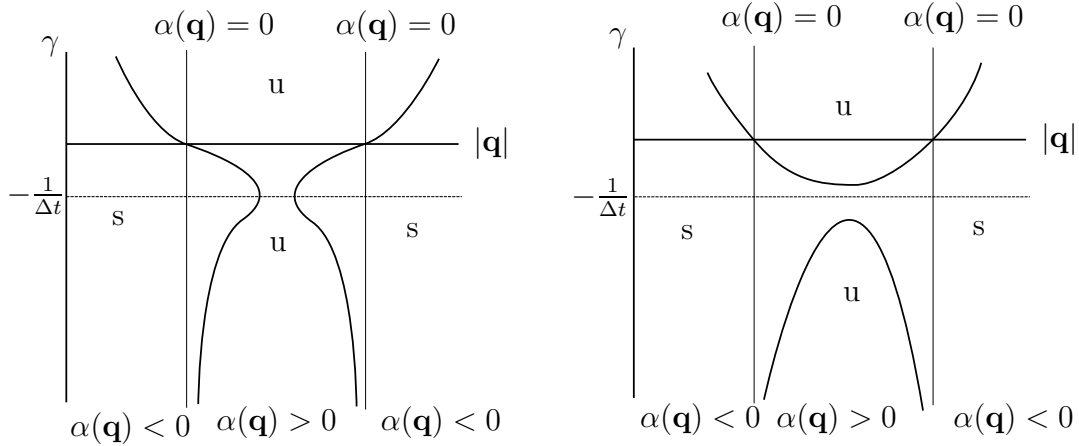, width=\textwidth}}
\end{center}
\caption{\label{fig:gammaq}Schematic diagram showing critical curves
   in unstable (left-hand picture) and stable (right-hand picture)
   cases. The transition between the two takes place at
   $\gamma=-1/\Delta t$. The regions of the diagram are marked as
   containing perturbations $\vec{q}$ which are stable (s) or unstable
   (u).}
\end{figure}
In Figure~\ref{fig:gammaq} we give a sketch of the shape of the
curves given by~\eqref{eq:cc1} and~\eqref{eq:cc2} for a fixed
direction of $\vec{q}$, in two cases, one before and one after a change  
of stability of the traveling waves. At points with
$\alpha(\vec{q})=0$, the curves either intersect $\gamma=0$ (if
$\nu\neq 2\pi n$), or asymptote to $\gamma=\pm\infty$ (when $\nu=2\pi
n$).

These curves divide regions of stable and unstable perturbations. For a traveling wave to be stable, it
must be stable for all perturbations $\vec{q}$.
One mechanism by which the stability of a traveling wave can change is
for two critical curves to collide at points with
$\frac{\partial\vec{q}}{\partial\gamma}=0$. Figure~\ref{fig:gammaq} shows a schematic of
this transition. Of course, the stability of the wave could also
hypothetically change without the two curves in the right hand picture
of Figure~\ref{fig:gammaq} colliding; the minimum of the top curve
would just have to extend below the maximum of the lower curve. We do
not consider this mechanism of stability change here, but it is
covered by the analysis in~\cite{MS04}.

Hence, we wish to identify points on the critical curves which have
$\frac{\partial\vec{q}}{\partial\gamma}=0$, since it is at these
points that the surfaces can collide and change the stability of the
traveling waves.
Differentiating~\eqref{eq:cc1} and~\eqref{eq:cc2} with respect to
$\gamma$, and setting $\frac{\partial\vec{q}}{\partial\gamma}=0$
gives:
\begin{align}
\gamma\sin\nu\frac{\partial\nu}{\partial\gamma}&=\cos\nu-1 \\
\Delta t\sin\nu&=\frac{\partial\nu}{\partial\gamma}(1-\gamma\Delta  
t\cos\nu)
\end{align}
Eliminating $\frac{\partial\nu}{\partial\gamma}$ results in
\[(\cos\nu-1)(1+\gamma\Delta t)=0.\]
When $\cos\nu=1$, the critical curves
asymptote to $\gamma=\pm\infty$, $\alpha(\vec{q})=0$, so the only
turning points with $\frac{\partial\vec{q}}{\partial\gamma}=0$ will be
at $\gamma=-\frac{1}{\Delta t}$, which is the value identified as optimal 
in~\cite{MS04}.

\section{Numerical results}
\label{sec:num}

We now give two numerical examples to show that there are regions of
parameter space in which traveling waves can be stabilized using the
methods described above. The first example we give can be stabilized
for $N=2$ but not, for the parameters we use, for $N=1$. The second example can be stabilized for
$N=1$. We also demonstrate the way in which stability of the traveling
wave is lost, as described above - two critical surfaces collide at
$\gamma=-1/\Delta t$.

We used the
Matlab package DDE-BIFTOOL~\cite{ddebiftool} to analyze the
linearized system~\eqref{eq:dde} and search for Hopf bifurcations in
order to locate the critical surfaces between stable and unstable
perturbations in $\gamma$-$\vec{q}$ space. For our first example, the
parameters used are $b_1=2.5$, $b_3=2$. Without loss of generality,
we choose $\vec{k}$ to be parallel to the $x$-axis. The spatial shifts
$\Dex{1}$ and $\Dex{2}$ are chosen to be parallel to the vector
$(1,1)$ and $\vec{k}$ respectively, and also to satisfy
$\Dex{j}\cdot\vec{k}=2\pi$, $j=1,2$. We choose $\rho_1=0.01$ and  
$\rho_2=0.007$. Recall that the
wave cannot be stabilized if there exist unstable perturbations with
$\vec{k}\cdot\vec{q}=0$ and $\vec{q}\cdot\Dex{j}=2\pi n_j$. In this
example, $q_{cr}=\frac{2(1-k^2)}{29}$ and it is simple to check in
each case that there are no
such $\vec{q}$ with $|\vec{q}|<q_{cr}$.

\begin{figure}
{\psfrag{qx}{\raisebox{-0.1cm}{$q_x$}}
\psfrag{qy}{$q_y$}
\subfigure[$N=0$]{\epsfig{file=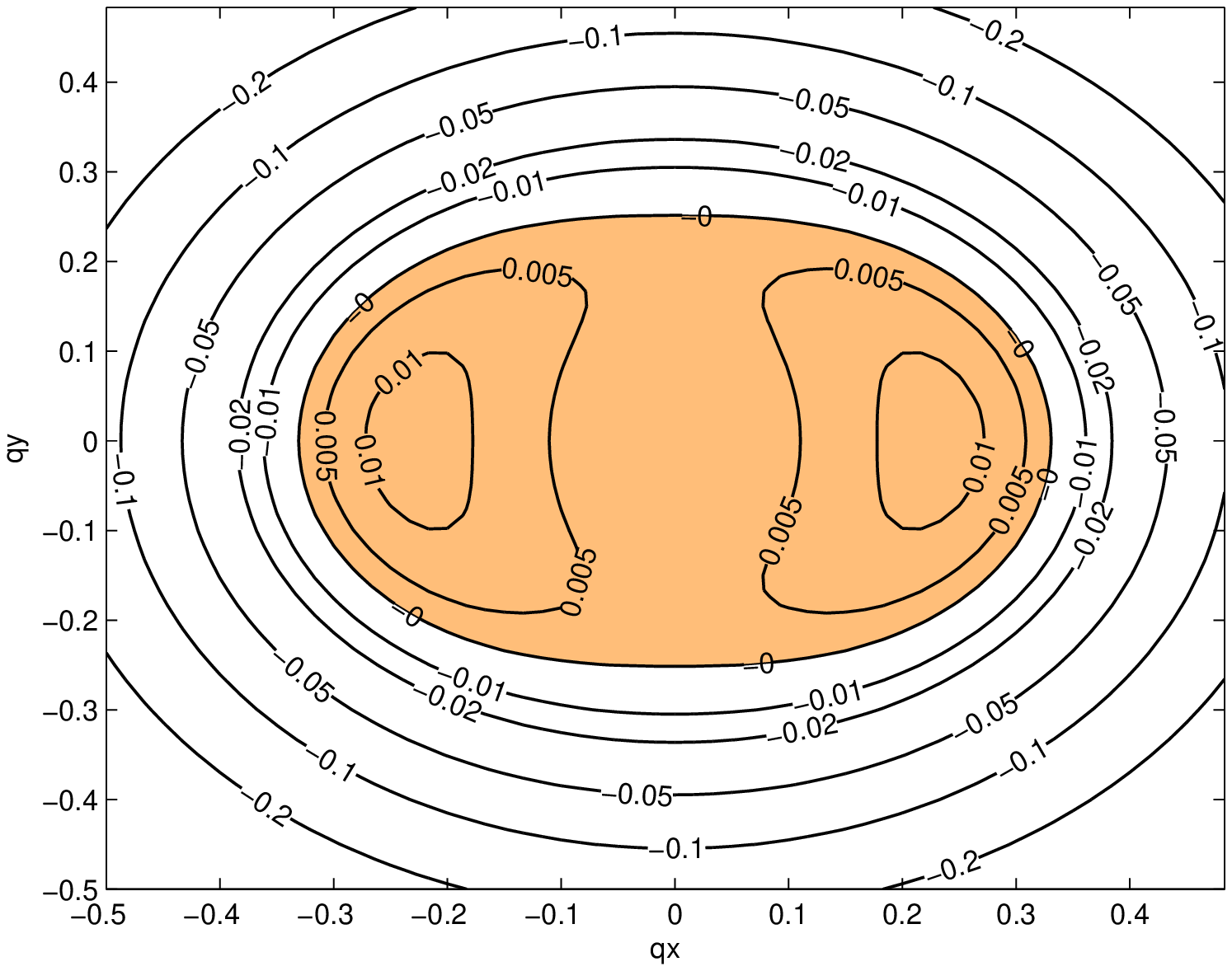,  
width=0.5\textwidth}}
\subfigure[$N=1$]{\epsfig{file=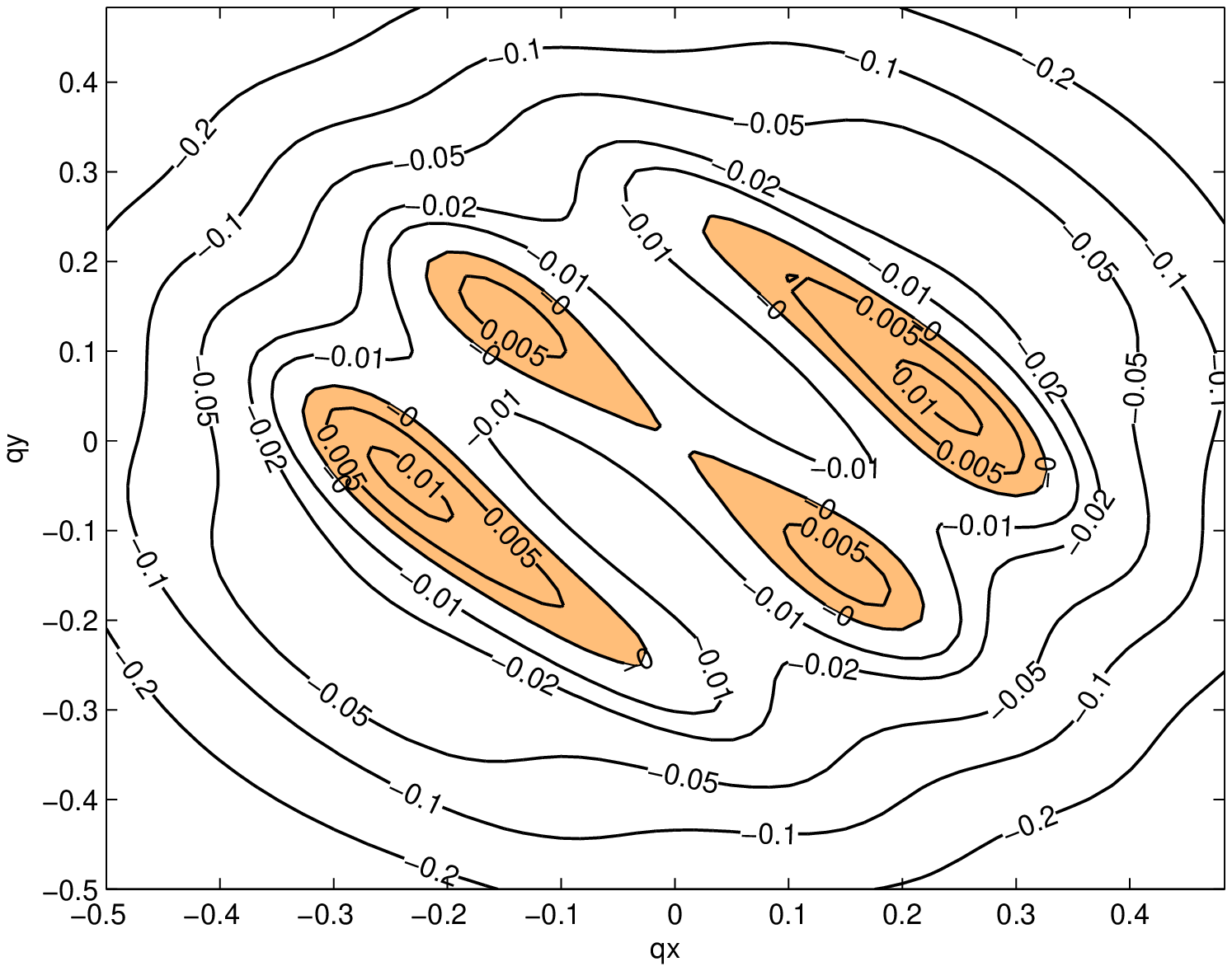,  
width=0.5\textwidth}}
\subfigure[$N=2$]{\epsfig{file=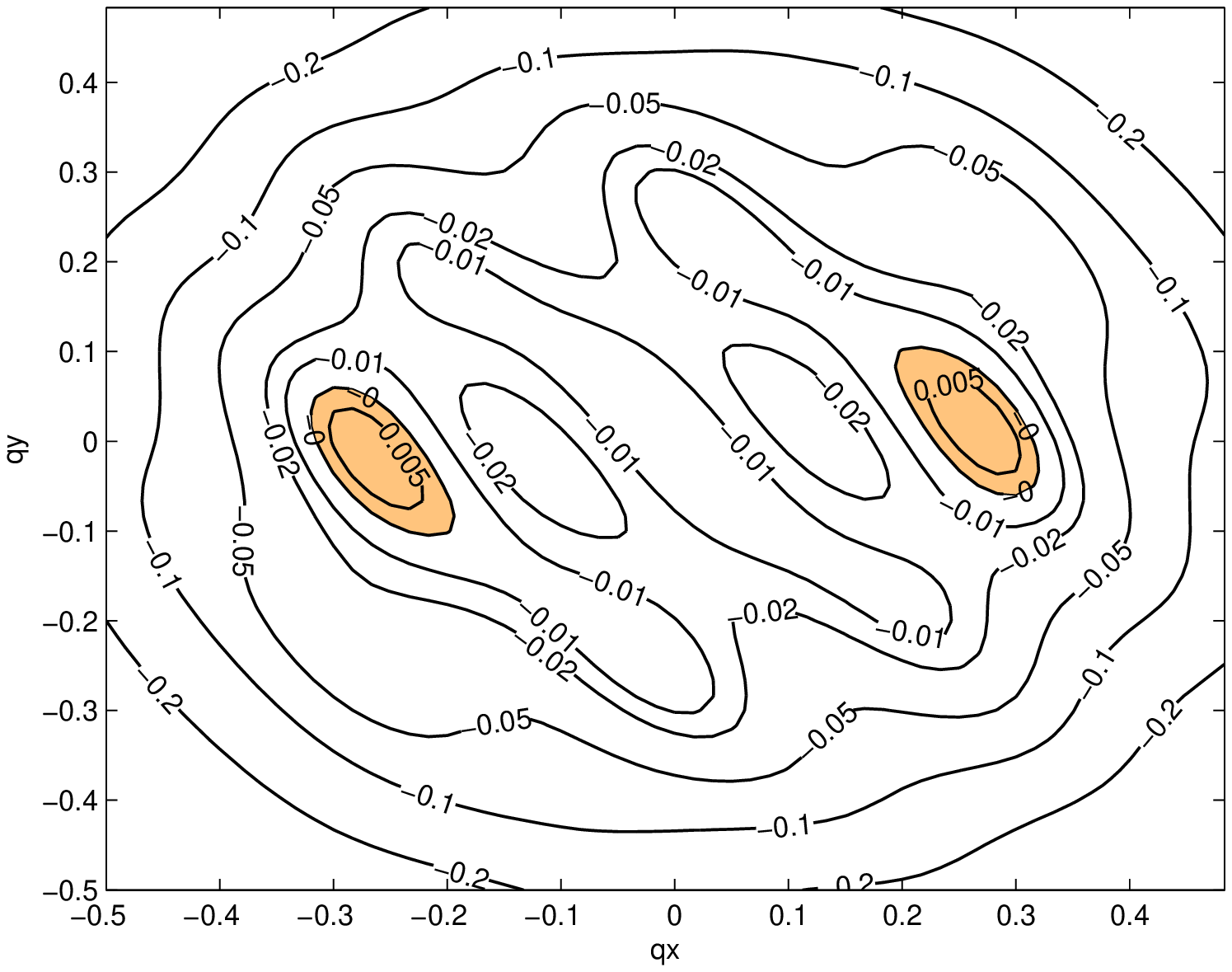,  
width=0.5\textwidth}}
\subfigure[$N=2$, $\rho_1=0.1$, $\rho_2=0.07$]{\epsfig{file=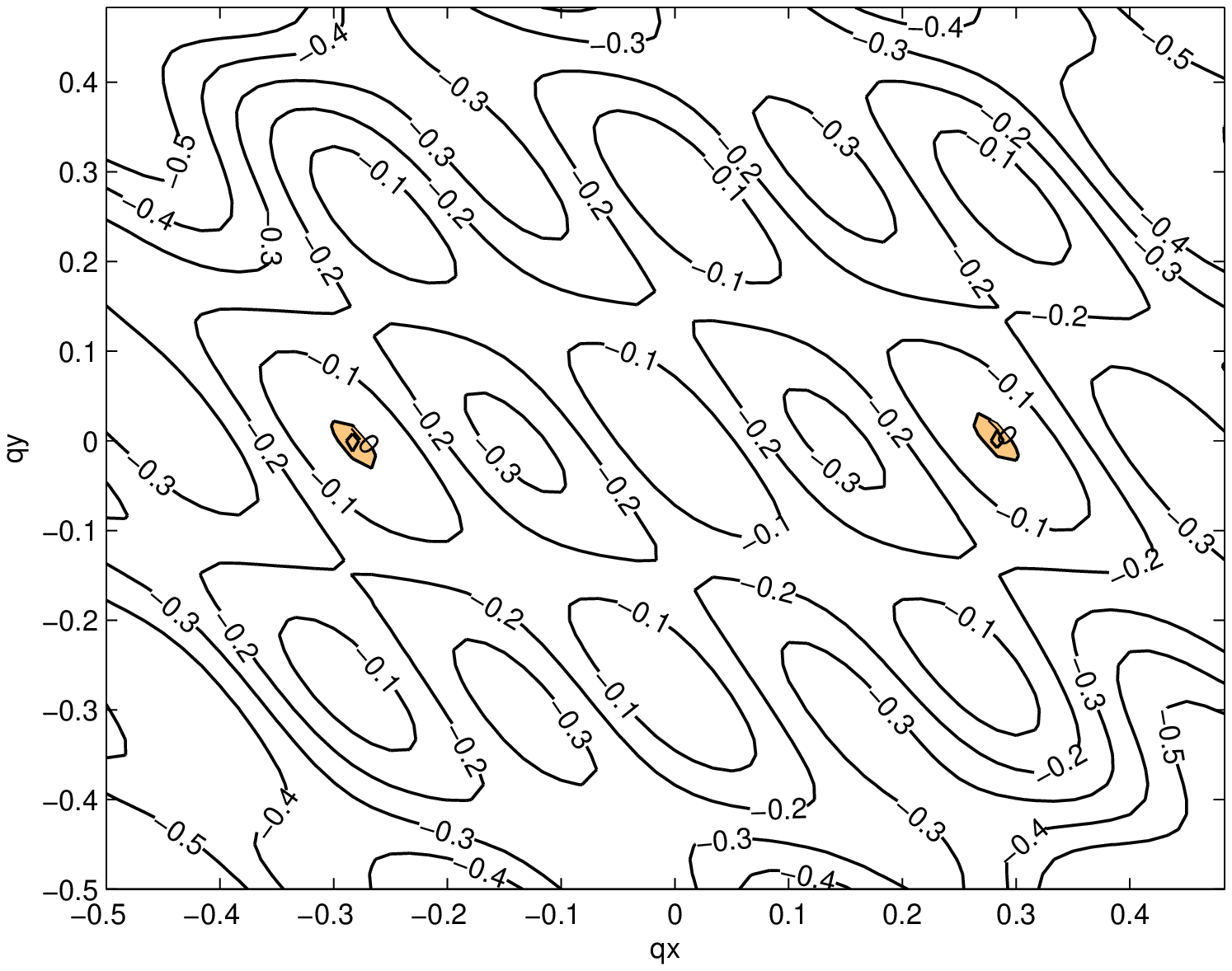,  
width=0.5\textwidth}}}
\caption{\label{fig:alpha}The figures show contour plots of
     $\alpha(\vec{q})$ close to $\vec{q}=\vec{0}$ in
     the three cases $N=0,1,2$. As $|\vec{q}|\rar\infty$,
     $\alpha(\vec{q})\rar-\infty$. Note that $\alpha(\vec{q})$  
corresponds to the largest real part of the
     eigenvalues of the linearized system with no temporal
     feedback (\ie $\gamma=0$). The regions which have positive  
eigenvalues (corresponding to instabilities) are shaded in.
Parameters used are $b_1=2.5$,
     $b_3=2$, and $\vec{k}=(0.285,0)$. $\Dex{1}$ is parallel to $(1,1)$,
     $\Dex{2}$ is parallel to $\vec{k}$, and
     $\Dex{j}\cdot\vec{k}=2\pi$. In (a), (b) and (c), $\rho_1=0.01$
     and $\rho_2=0.007$, which are the parameters used in later
     investigations. In (d) $\rho_1$ and $\rho_2$ are increased by a
     factor of ten, and there are still unstable regions, which cannot
     be made to vanish no matter how large we choose the $\rho_j$. 
 The plots of $\alpha(\vec{q})$ for the
     other values of $\vec{k}$ used in Section~\ref{sec:num} are very
     similar, and so we do not show them here.}
\end{figure}
The contour plots of $\alpha(\vec{q})$ in Figure~\ref{fig:alpha} show  
the regions of
$\vec{q}$-space in which it is possible to find instabilities - that
is, those regions where $\alpha(\vec{q})>0$. Note that there is a symmetry
$\vec{q}\rar-\vec{q}$. In the case $N=1$
there are two disjoint regions, but for $N=2$ there is only one. If
there were no temporal feedback (\ie $\gamma=0$), then the
traveling wave would be unstable. In this example it is not possible
to suppress all the instabilities using just the spatial feedback;
figure~\ref{fig:alpha}(d) has spatial feedback ten times that of
(a)-(c). The $\vec{q}$ which are still unstable are centered around
points which satisfy
$\vec{q}\cdot\Dex{j}=2\pi m$ for some $m\in\N$, so the terms
proportional to $\rho_j$ in~\eqref{eq:J} vanish (\ie these $\vec{q}$
are unaffected by the spatial feedback). Therefore, however
large the $\rho_i$ are in this example, these $\vec{q}$ will still be unstable.
 It may, in other examples, be possible 
that the traveling wave can be stabilized using only
spatial feedback. Examples of this for the one--dimensional CGLE 
are shown in~\cite{MS04}. We now give more details on the addition of
temporal feedback to the two cases $N=1$ and $N=2$.

\subsection{N=2}

Parameters are chosen as described above. We consider two values of
$\vec{k}$, $\vec{k}=(0.285,0)$ and $\vec{k}=(0.287,0)$. In both cases,
the value of $-1/\Delta t\approx -0.04$.

\begin{figure}
{\psfrag{qx}{$q_x$}
\psfrag{qy}{$q_y$}
\psfrag{qy;}{$q_y$}
\psfrag{g}{$\gamma$}
\subfigure[$|\vec{k}|=0.285$]{\epsfig{file=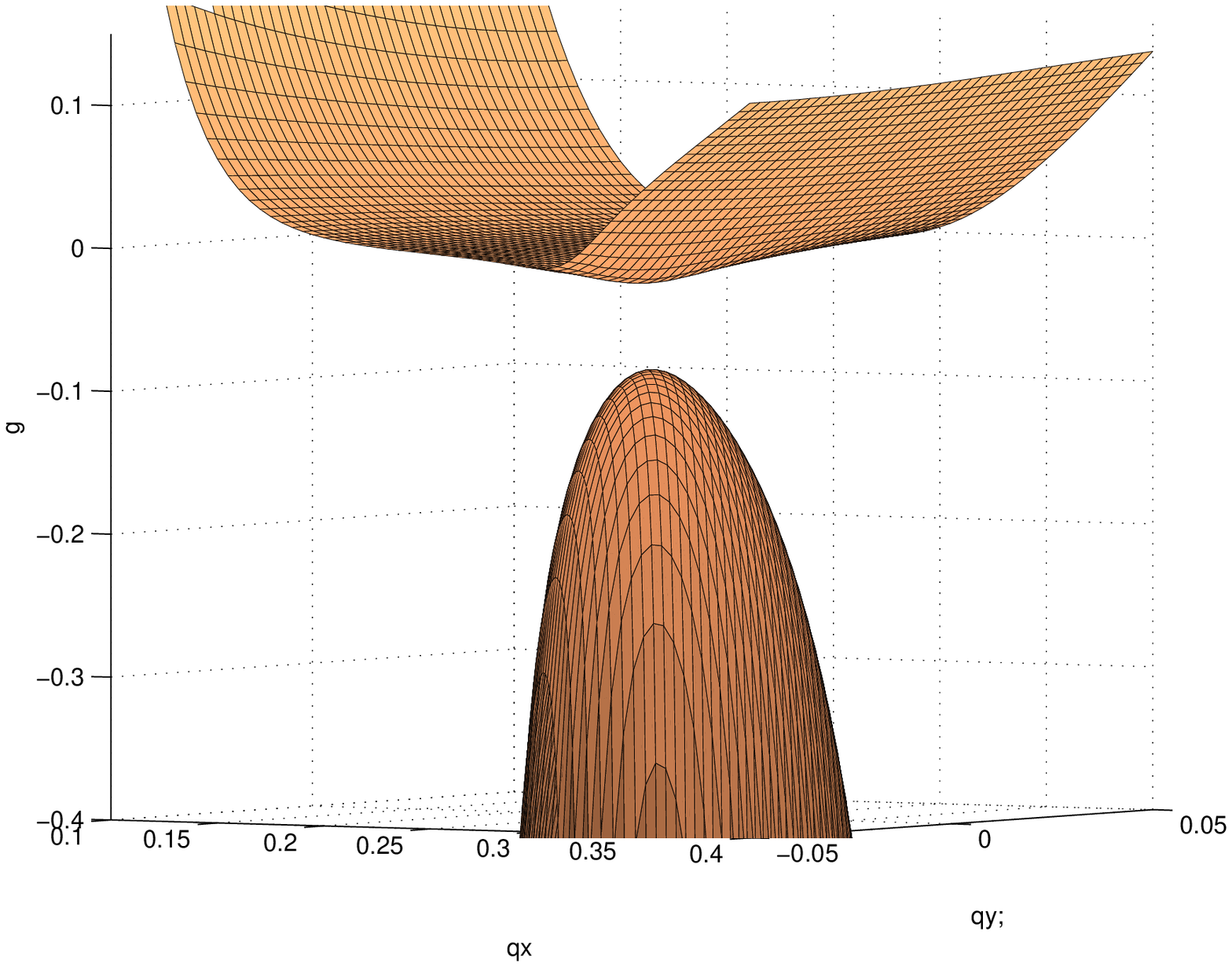, 
width=0.5\textwidth}}
\subfigure[$|\vec{k}|=0.287$]{\epsfig{file=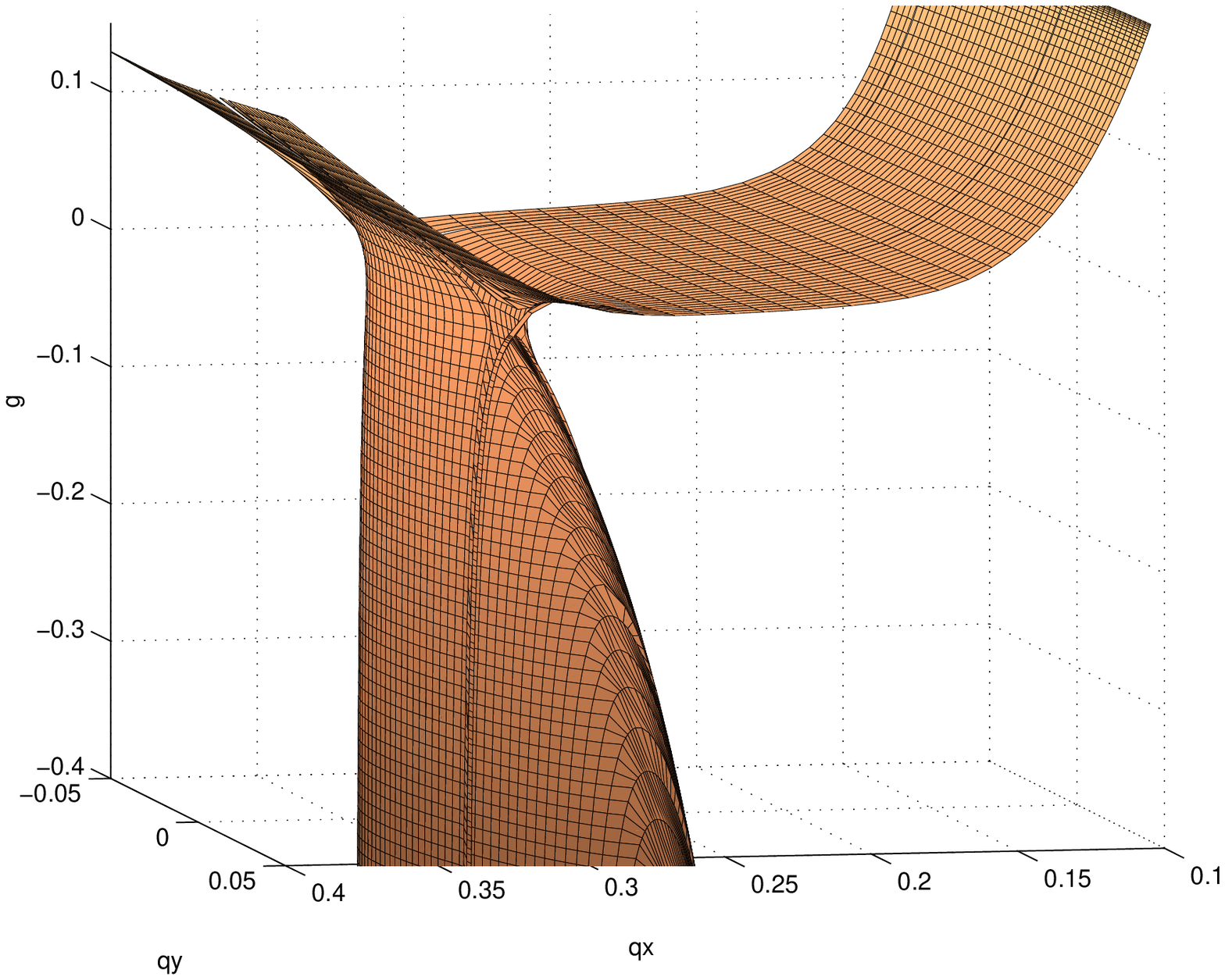, 
width=0.5\textwidth}}}
\caption{\label{fig:hopf_surfsN2} The two plots show surfaces of
   Hopf bifurcations for two values of $|\vec{k}|$. The regions above the  
upper surface and below the lower surfaces are regions of unstable  
perturbations. In (a) the surfaces
   are disjoint and so there is some range of $\gamma$ for which all  
perturbations will decay and so the
   traveling waves can be stabilized. In (b) this is not the
   case. Notice that the range of $\vec{q}$ in these plots includes
that where $\alpha(\vec{q})> 0$ (see
   figure~\ref{fig:alpha}(c)). Compare also with
   figure~\ref{fig:fm_N2}, which shows plots of the Floquet
   multipliers at $\gamma=-1/\Delta t$. Parameters used are $b_1=2.5$, $b_3=2$,
   $\rho_1=0.01$, $\rho_2=0.007$. $\Dex{1}$ is parallel to $(1,0)$ and
   $\Dex{2}$ is parallel to $(1,1)$.
}
\end{figure}

Figure~\ref{fig:hopf_surfsN2} shows bifurcation surfaces for the two
  cases. In the first case, the surfaces are disjoint and
so the traveling wave can be stabilized for a range of choices of
$\gamma$, which include $-1/\Delta t$. In the second case, the two
surfaces have joined, and there is no choice of $\gamma$ for which all
$\vec{q}$ perturbations are stable.

\begin{figure}
{\psfrag{qx}{\raisebox{-0.1cm}{$q_x$}}
\psfrag{qy}{$q_y$}
\subfigure[$|\vec{k}|=0.285$,  
$N=2$]{\epsfig{file=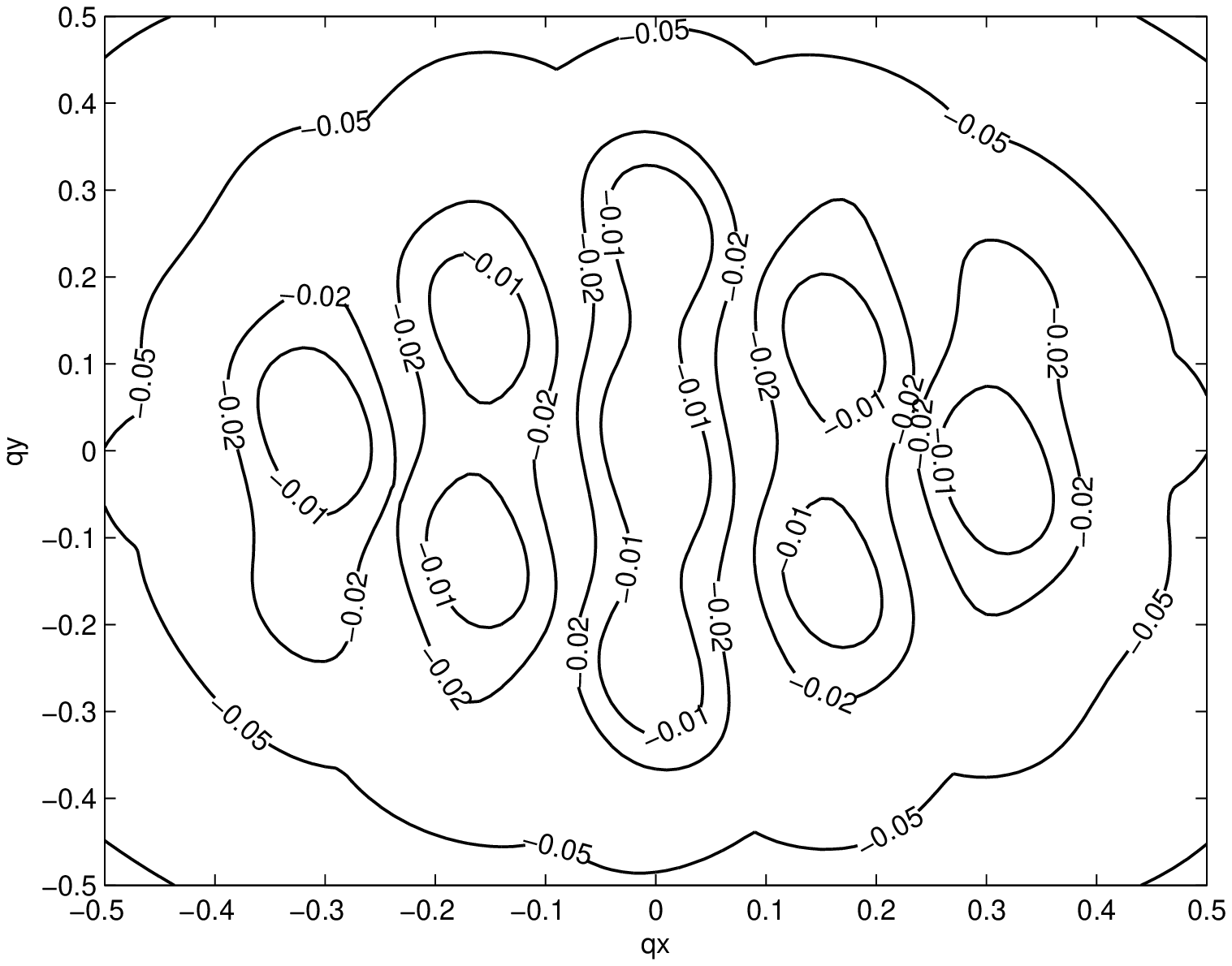, width=0.5\textwidth}}
\subfigure[$|\vec{k}|=0.287$,  
$N=2$]{\epsfig{file=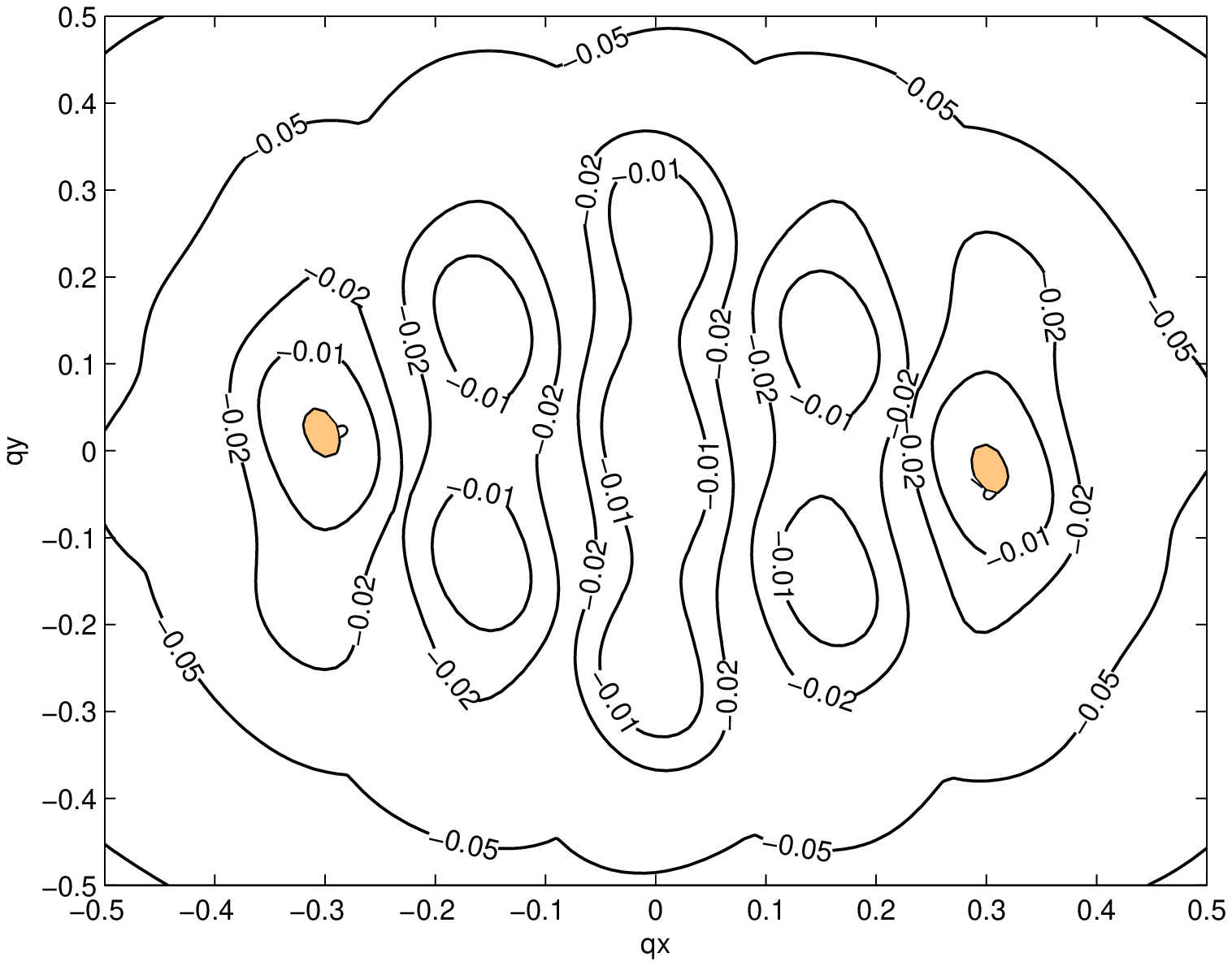, width=0.5\textwidth}}
\subfigure[$|\vec{k}|=0.283$, $N=1$]{\epsfig{file=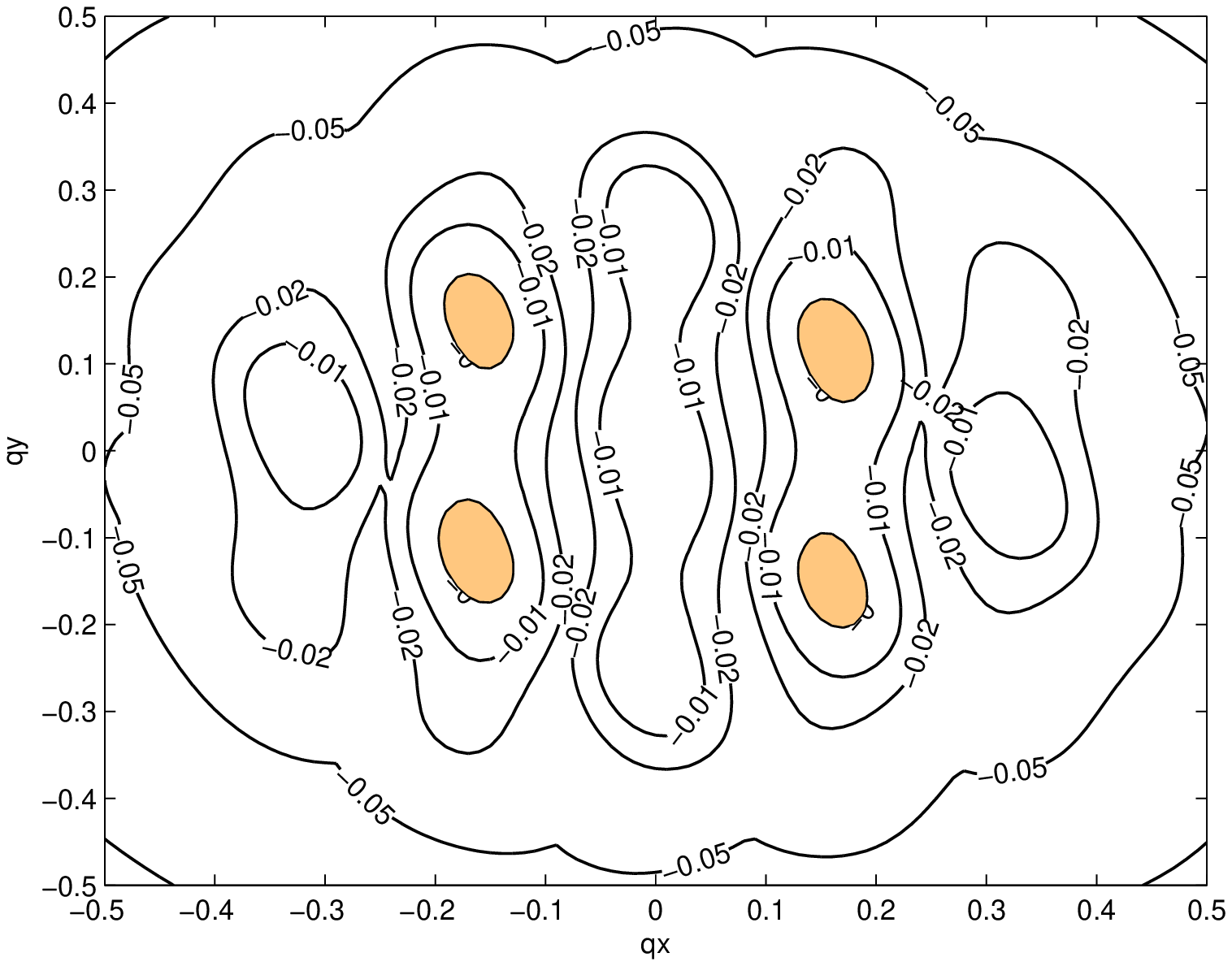,  
width=0.5\textwidth}}
\subfigure[$|\vec{k}|=0.285$, $N=1$]{\epsfig{file=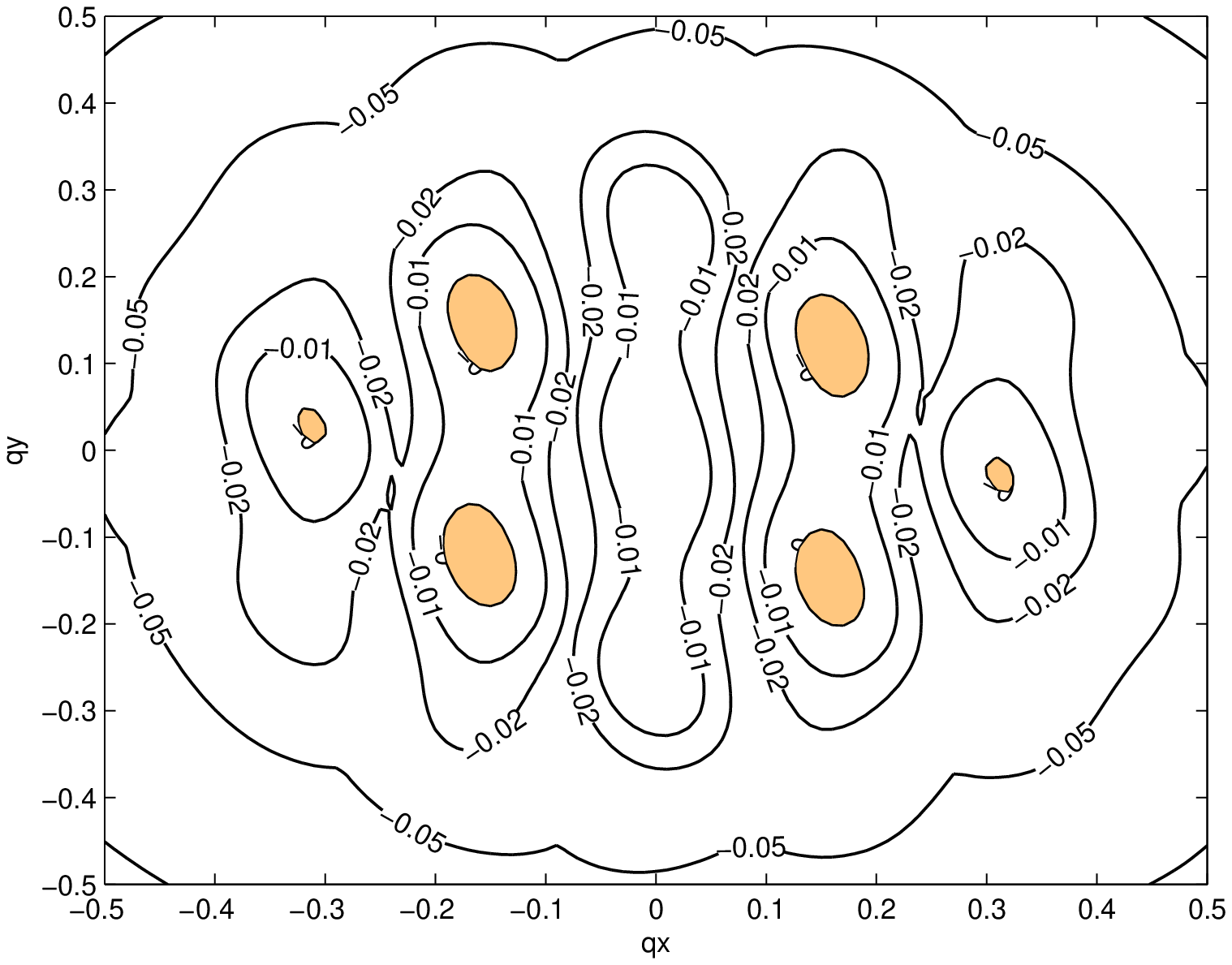,  
width=0.5\textwidth}}}
\caption{\label{fig:fm_N1} \label{fig:fm_N2} The contour plots show the  
real parts of the Floquet exponents
   of the linearized system at $\gamma=-1/\Delta t\approx-0.04$. The  
upper two plots show $|\vec{k}|=0.285,0.287$, $N=2$, and the lower two  
show
   $|\vec{k}|=0.283,0.285$, $N=1$. The regions with positive growth rate 
are
   shaded in, as in Figure~\ref{fig:alpha}. We can see that for $N=2$,  
the
   case $|\vec{k}|=0.285$ is stable but the case $|\vec{k}|=0.287$ is  
unstable. For $N=1$ there are additional regions of instability so the  
wave is not stabilized. Parameters used are $b_1=2.5$, $b_3=2$,
   $\rho_1=0.01$, $\rho_2=0.007$. $\Dex{1}$ is parallel to $(1,0)$ and
   $\Dex{2}$ is parallel to $(1,1)$.
}
\end{figure}

Figure~\ref{fig:fm_N2} shows the real part of the Floquet exponents of  
the linearized
system at $\gamma=-1/\Delta t$. We can see that for $|\vec{k}|=0.285$,
the growth rates are negative for all $\vec{q}$, so the
traveling wave is stable. For $|\vec{k}|=0.287$, there are two (symmetry-related)
regions in the $\vec{q}$-plane which have positive growth rates
and hence the traveling wave is unstable.

\subsection{N=1}

We now investigate whether the traveling waves can be stabilized
using only one spatially shifted term. We first investigate the same  
example as used above
for $N=2$, but only use the spatial shift $\Dex{1}$ which is parallel
to the vector $(1,1)$. (Note that we cannot use only the second
spatial shift $\Dex{2}$ since it is parallel to
$\vec{k}$.)
Figure~\ref{fig:fm_N1} shows the real part of the Floquet exponents of  
the
linearized system at $\gamma=-1/\Delta t$ for this case, for two
values of $|\vec{k}|$. Note that there are more regions of instability  
in the $N=1$ case than the $N=2$ case. The instability at  
$\vec{q}\approx (0.3,0)$ is present in both cases, and is suppressed as  
$|\vec{k}|$ is varied. However, in the $N=1$ case, there are additional  
regions of unstable $\vec{q}$ which are not stabilized at these  
parameter values.

\begin{figure}
{\psfrag{qx}{\raisebox{-0.1cm}{$q_x$}}
\psfrag{qy}{$q_y$}
\subfigure[\label{fig:N1staba}  
$\gamma=0$]{\epsfig{file=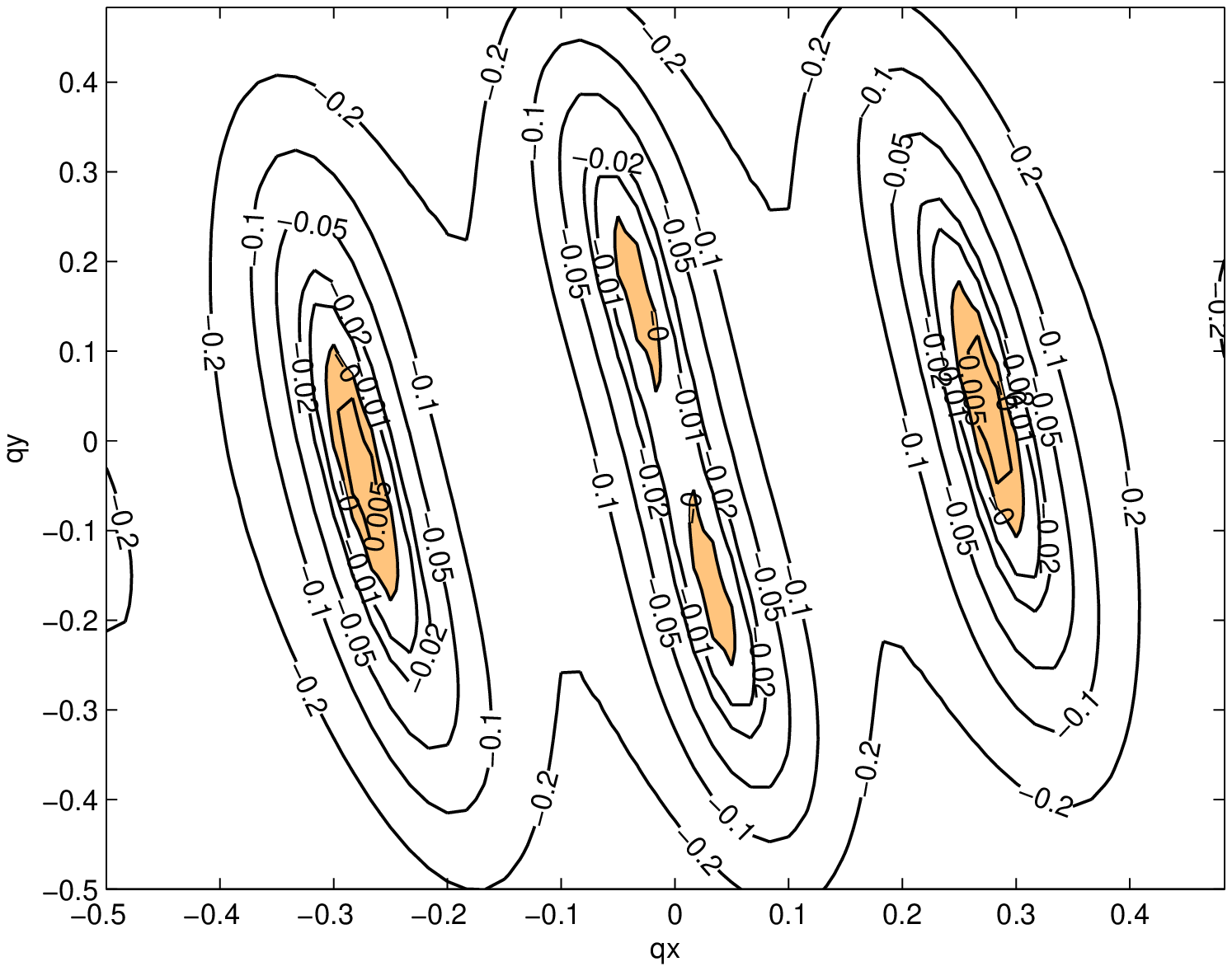, width=0.5\textwidth}}
\subfigure[\label{fig:N1stabb} $\gamma=-1/\Delta t$]  
{\epsfig{file=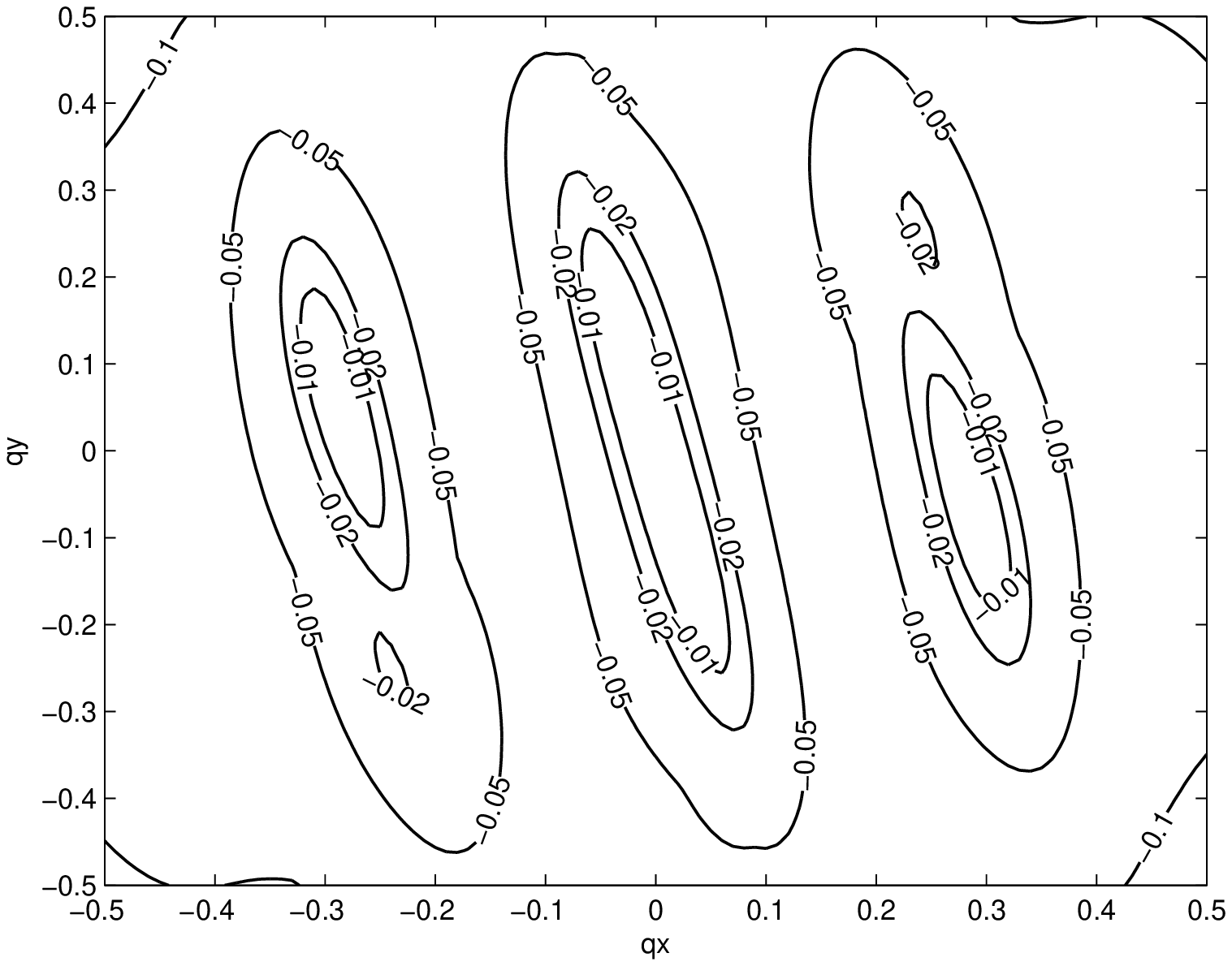, width=0.5\textwidth}}}
\caption{\label{fig:fm_N1_stab} The two plots show the real part of the  
Floquet exponents
   of the linearized system for $|\vec{k}|=0.284$, and $N=1$.  
In~\ref{fig:N1staba}
   there is no temporal feedback (\ie $\gamma=0$). In~\ref{fig:N1stabb}
   $\gamma=-1/\Delta t$. The regions with positive growth rate are
   shaded in, as in Figure~\ref{fig:alpha}, so we can see that with the
   temporal feedback, the plane wave is stabilized. Parameters used are $b_1=2.5$, $b_3=2$,
   $\rho_1=0.1$, and $\Dex{1}$ is parallel to $(1,0.2)$. 
}
\end{figure}
Figure~\ref{fig:fm_N1_stab} shows the real part of the Floquet  
exponents for a second
parameter set, for which the wave can be stabilized using only one  
spatially shifted term. In this second example, all parameters are the  
same except that the spatial shift
$\Dex{1}$ is parallel to the vector $(1,0.2)$, and $\rho_1=0.1$.  The  
two plots
show the real parts of the 
exponents of the linearized system with and without the
temporal feedback (that is, $\gamma=0$ and $\gamma=-1/\Delta
t$). There are no positive growth rates  when $\gamma=-1/\Delta t$ and
hence the traveling wave can be stabilized.

\section{Discussion}
\label{sec:conc}

In this paper we have examined the possibility of stabilizing
traveling waves of the CGLE in two spatial dimensions using a
combination of both temporal and spatial feedback. The feedback is
noninvasive, in that it vanishes at the targeted wave solution, and is
a generalization of that first proposed by Pyragas~\cite{Pyr92}. Our
analysis is a local linear stability analysis which involves the study
of a linear delay equation. The analysis is similar to that
in~\cite{MS04}, for the 1D CGLE. However, in the 2D case, the spatial
shifts $\Dex{j}$ must be carefully chosen as otherwise there may be
some perturbations $\vec{q}$ which are unaffected by the spatial
feedback, meaning the wave cannot be stabilized for any values of the
control parameters. We show that if the $\Dex{j}$ are correctly
chosen, the results of~\cite{MS04} that the `best' value of the gain is
$\gamma=-1/\Delta t$ still holds.

We expect that this method of using noninvasive feedback to stabilize
otherwise unstable waves could also be applied to amplitude equations
other than the CGLE. One property of the CGLE which simplifies the
analysis is that the form of the wave and its dispersion relation can
be written down in closed form, and are particularly simple. For other
amplitude equations this will not be the case. A lack of knowledge of
the dispersion relation means that it is not possible to choose {\it a
priori} the time delay and two spatial shifts in a consistent way so
that all temporal and spatial feedback terms vanish simultaneously.
It may be possible, however, to circumvent this difficulty if just one
spatial term and one temporal delay term are employed in the feedback;
a single wave may then satisfy the criterion that the feedback vanish
when it is realized. However, the unknown dispersion relation will
then be selecting its direction of travel.

In our analysis, we assumed the feedback parameters $\gamma$ and
$\rho_j$ were real. This is a
mathematically convenient choice since the control matrix then ends up
being a multiple of the identity. However, the choice of
{\it real} gain coefficients for the control terms does not seem
natural when all other coefficients in the {\it complex}
Ginzburg--Landau equation are complex. Future work on this problem
will consider the case of complex gain parameters.

\section*{Acknowledgements}

The authors would like to thank Luis Mier y Teran for assistance using
DDE-BIFTOOL. This research was supported by NSF grant DMS-0309667.

\end{document}